\documentclass[prb,twocolumn,showpacs,preprintnumbers,amsmath,amssymb]{revtex4}
\usepackage{graphicx}
\usepackage{dcolumn}
\usepackage{bm}

\begin{document}

%\preprint{APS}
\title{Electronic structure of spin 1/2 Heisenberg antiferromagnetic
systems: Ba$_2$Cu(PO$_4$)$_2$ and Sr$_2$Cu(PO$_4$)$_2$\\
}
\author{Sarita S. Salunke} 
\affiliation{Department of Physics, Indian Institute of Technology Bombay, Mumbai 400076, India.}
\author{M. A. H. Ahsan}
\altaffiliation{Present address: Jamia Millia Islamia, New Delhi-110025, India}
\author{R. Nath}
\altaffiliation{Present address: Max-Planck-Institute for Chemical Physics of Solids, N\"othnitzer Str. 40 01187 Dresden, Germany}
\author{A.V. Mahajan} 
\author{I. Dasgupta}
\email{dasgupta@phy.iitb.ac.in}
\affiliation{Department of Physics, Indian Institute of Technology Bombay, Mumbai 400076, India.}

\date{\today}
\begin{abstract}
We have employed first principles calculations
to study the electronic structure and magnetic properties of the low-
dimensional phosphates, Ba$_2$Cu(PO$_4$)$_2$ and Sr$_2$Cu(PO$_4$)$_2$. Using
the self-consistent tight binding linearized muffin-tin orbital (TB-LMTO)
method and the N$^{th}$ order muffin tin orbital 
(NMTO) method we have calculated
the various intrachain as well as the interchain hopping parameters 
between the magnetic ions (Cu$^{2+}$) for 
both the compounds. We 
find that the nearest-neighbor intrachain hopping ($t$) is the dominant 
interaction 
suggesting  the compounds to be indeed one-dimensional. Our analysis of the
band dispersion, orbital projected band structures and the hopping
parameters confirms that the Cu$^{2+}$ - Cu$^{2+}$ super-super exchange
interaction takes place along the crystallographic $\it{b}$-direction mediated by
O-P-O. We have also analyzed in details the origin of short range exchange 
interaction for these systems. Our {\em ab-initio} 
estimate of the ratio of the 
exchange interaction of Sr$_2$Cu(PO$%
_4$)$_2$ to that of Ba$_2$Cu(PO$_4$)$_2$ compares excellently with available
experimental results.
\end{abstract}

\maketitle



\section{\label{sec:level1}Introduction}

Low-dimensional quantum spin systems with chain, ladder, or planar
geometries have attracted much attention due to their unconventional
magnetic properties. \cite{lem} Much effort has been devoted particularly
over the last decade to understand the behavior of quasi-one-dimensional
spin systems. These systems exhibit a rich variety of phases due to the
enhanced quantum fluctuations in reduced dimensionality. It is well known 
that the half-integer uniform antiferromagnetic chain, though
having a singlet ground state, is gapless and quantum critical 
with power law spin correlations.  
A deviation from the uniform chain due to, for instance, an
alternating exchange leads to a gapped (spin-gap $\Delta $)ground state. 
Frustration arising in one dimensional spin chains due to next 
nearest neighbor interactions can also lead to a gap in the spin excitation spectrum.
A well known case is that of the Majumdar-Ghosh(MG) chain \cite{ckmdkg} 
where the Hamiltonian is antiferromagnetic in nature, 
and includes both nearest neighbor $J_{1}$ and next nearest neighbor 
$J_{2}$ exchange interactions. This model is exactly solvable 
for the ratio $\frac{J_{2}}{J_{1}}$ = $\alpha$ = 0.5. The MG-chain 
supports a spin gap for $\alpha$ $>$ $\alpha_{cr}$ = 0.2411. For 
$0 < \alpha < \alpha_{cr}$, a gapless phase is obtained.
In contrast, integer-spin
uniform AF chains are always gapped as conjectured by Haldane \cite{haldane}
as also half-integer even-leg ladders \cite{dagottorice}
while the half-integer odd-leg ladders are gapless.
 Recently, the
discovery of spin-Peierls transition in the one-dimensional (1D) Heisenberg
antiferromagnet (HAF) CuGeO$_{3}$ \cite{hase} renewed interest in 1-D
transition metal oxides with spin $S=1/2$ ions such as Cu, which are
responsible for both the magnetic properties as well as the Jahn-Teller
distortion of the lattice.

Low-dimensional quantum spin systems are important also due to the belief that understanding these systems may be crucial to understand the physics of
high-T$_{c}$ cuprates. \cite{dagatto_koti}
In addition, the field of low-dimensional quantum
magnetism provides a fertile ground for rigorous theory. Powerful techniques
like the Bethe Ansatz (BA) and bosonization are available to study ground
and excited state properties of these systems. Models of interacting spin
systems are known for which the ground state and in some cases their
low-lying excitation spectra are known exactly. The knowledge gained from
these models provides impetus to look for real materials so that
experimental confirmation of the theoretical predictions can be made.

Quasi-1D magnets, in which the magnetic interaction in one direction
dominates, with much weaker interactions in other directions, exhibit
short-range order over a wide temperature range. A uniform Heisenberg
half-integer-spin chain as discussed above 
has a gapless spin excitation spectrum. The ground
state does not have long range order (LRO) because of strong quantum fluctuations. In real
systems however, inter-chain interactions can lead to a deviation from ideal
behavior and even magnetic LRO. Whereas a large number of quasi-1D systems have 
been studied, there are only a few cases in which the interchain interactions
are very weak and true 1D behaviour can be observed down to low-temperatures.
One example is Sr$_{2}$CuO$_{3}$ where the exchange
coupling constant $J$/$k_{B}$ is about $2200\pm 200$ K and the ordering
temperature $T_{N}$ $=5.4$ K, giving the ratio $k_{B}T_{N}$/$J\approx $ $%
0.25\%$. \cite{takano} Another example is copper benzoate Cu(C$_{6}$H$_{5}$%
COO)$_{2}$$\cdot$3H$_{2}$O with $J$/$k_{B}$ $=17.2$ K. \cite{asano} 

Very recently, the susceptibility measurements by Belik \textit{et al.} \cite%
{belik} and susceptibility and nuclear magnetic resonance (NMR) measurements
by Nath \textit{et al.} \cite{nath} have suggested that Ba$_{2}$Cu(PO$_{4}$)$%
_{2}$ ($J/k_{B}=151$ K) and Sr$_{2}$Cu(PO$_{4}$)$_{2}$ ($J/k_{B}=165$ K)
are excellent $S=1/2$ linear chain Heisenberg antiferromagnets. The
temperature dependence of the NMR shift $K(T)$ is well described by the $%
S=1/2$ Heisenberg antiferromagnetic chain model. \cite{john} The $J/k_{B}$
in these compounds is about one order of magnitude smaller than in Sr$_{2}$%
CuO$_{3}$ and about one order of magnitude larger than in copper benzoate.
In Sr$_{2}$Cu(PO$_{4}$)$_{2}$, Belik \textit{et al.} \cite{belik} suggested
from their measurements that LRO sets in at $T_{N}=0.085$ K, which implies
that  $k_{B}$$T_{N}/J$ $=$ $0.06\%$, making Sr$_{2}$Cu(PO$_{4}$)$_{2}$ 
\textit{the best} realization of a 1D $S=\frac{1}{2}$ HAF system studied to date. \
This provides a rare opportunity for a comparison of experimental results
with theoretical models and improving our understanding on these systems.

The unique 1D magnetic behavior exhibited by these compounds may be traced
back to the interplay of the geometry (crystal structure) and quantum
chemistry and therefore requires a thorough understanding of the electronic
structure of these systems. In this paper, we shall study the electronic
structure of Ba$_{2}$Cu(PO$_{4}$)$_{2}$ and Sr$_{2}$Cu(PO$_{4}$)$_{2}$ in
some detail. The characteristic feature of these compounds is a chain of isolated CuO$%
_{4}$ square plaquettes along the crystallographic $\mathit{b}$
direction. The goal of the present work is to understand and evaluate the
various exchange paths leading to 1D magnetism by calculating the various
hopping integrals. The latter will also provide the dominant exchange
integrals entering the magnetic Hamiltonian for the spin-half Heisenberg
chain given by 
\begin{equation}
H=-\sum_{ij}J_{ij}(\bf{S}_{i}\cdot \bf{S}_{j})
\end{equation}%
where $J_{ij}$ is the magnetic exchange interaction and $S$ denotes the spin. The various hopping matrix elements
between the
magnetic ions (Cu for these compounds) are strongly dependent on the way the
CuO$_{4}$ squares are assembled. The edge sharing CuO$_{4}$ squares and
corner sharing CuO$_{4}$ squares lead to quite different values for the
nearest-neighbor hopping ($t$) and the exchange interaction ($J$). 
In the latter case, the 180%
$^{\circ }$ Cu-O-Cu bond with a common O 2$p$ orbital gives rise to large
magnitude of $J$ as seen for the CuO$_{2}$ planes in the high-T$_{c}$
cuprates. This contrasts with the 90$^{\circ }$ Cu-O-Cu bond when two CuO$%
_{4}$ squares share an edge, where the hopping matrix element $t$ and
therefore the exchange interaction are expected to be small in magnitude. In
the present work, we shall explicitly obtain the various effective Cu-Cu
hopping parameters using N-th order muffin-tin orbital method (NMTO) and
clarify the role of electronic structure on the magnetic properties of these
systems.

The remainder of the paper is organized as follows:  In Sec. II, we
describe the crystal structure of both Ba$_2$Cu(PO$_4$)$_2$ and Sr$_2$Cu(PO$%
_4$)$_2$ followed by the computational details. Section III is devoted to the
detailed discussion of the electronic structure using tight-binding
linearized muffin-tin orbital method (TB-LMTO) within atomic sphere
approximation (ASA) and the NMTO method followed by a comparative
study of our calculations with available experimental results. Finally, in
Sec. IV, we present our conclusions.

\section{Crystal Structure and Computational details}

Ba$_{2}$Cu(PO$_{4}$)$_{2}$ and Sr$_{2}$Cu(PO$_{4}$)$_{2}$ are two
isostructural compounds in the family of $S=1/2$ 1D HAF systems. \cite%
{structure} They both crystallize in the monoclinic structure with space
group C$_{2/m}$ (No. 12). The lattice parameters are $\mathit{a}$ = 12.16 
\AA , $\mathit{b}$ = 5.13 \AA , $\mathit{c}$ = 6.88 \AA, and $\mathit{\beta }$
= 105.42$^{\circ }$ and $\mathit{a}$ = 11.51 \AA , $\mathit{b}$ = 5.07 \AA , 
$\mathit{c}$ = 6.57 \AA, and $\mathit{\beta }$ = 106.35$^{\circ }$ for Ba$_{2}
$Cu(PO$_{4}$)$_{2}$ and Sr$_{2}$Cu(PO$_{4}$)$_{2}$, respectively. The
characteristic features of the structure as illustrated in Fig. 1 are
isolated, square, CuO$_{4}$ plaquettes sharing their edges with two similar
kind of PO$_{4}$ tetrahedra. This edge sharing via PO$_{4}$ tetrahedra takes
place along one crystallographic direction ($\mathit{b}$ direction) forming
isolated [Cu(PO$_{4}$)$_{2}$]$_{\infty}$ chains along the crystallographic
$\mathit{b}$ direction. The chains, are however, staggered with
respect to each other along the crystallographic $\mathit{a}$ direction.
The Ba/Sr
cations reside between these parallel chains. In Fig. 1(b), we have indicated 
the hoppings between various Cu atoms. In Fig. 1(c), we have displayed
the schematic energy level diagram due to crystal field splitting of the Cu $d$ ion in
the square planar coordination provided by the oxygens. We find that the
highest energy level is Cu $d_{x^{2}-y^{2}}$ and is well separated from
the rest of  Cu $d$ levels.

\begin{figure}[tbp]
\includegraphics{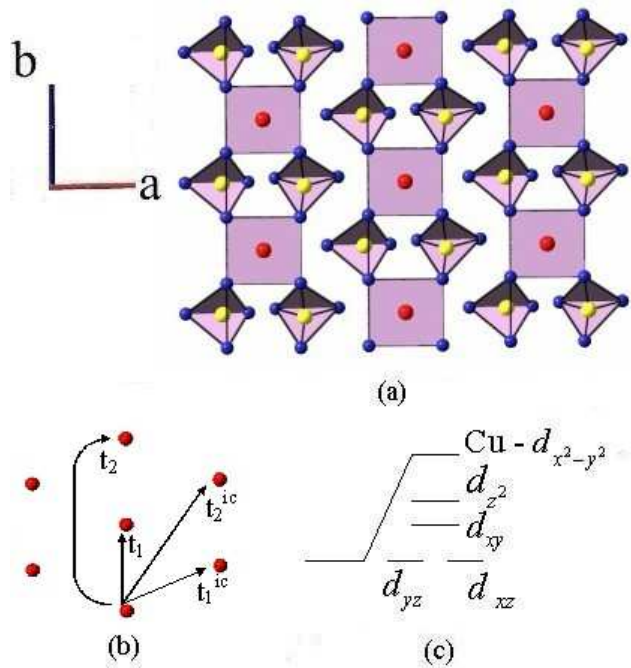}% Here is how to import EPS art
\caption{(Color online) 
(a) Schematic diagram of [Cu(PO$_{4}$)$_{2}$]$_{\infty }$ linear chains
propagating along the \textit{b} direction for (Ba/Sr)$_{2}$Cu(PO$_{4}$)$_{2}$. (b) Various hoppings between the Cu atoms.
(c) Schematic energy level diagram for a Cu $d$ ion in square planar environment }
\label{fig:structure}
\end{figure}
Our analysis of the electronic structure is carried out in the framework of
the (TB-LMTO) method \cite{lmto,lmtocode} in the (ASA) within local density approximation (LDA)
to the density functional theory. The space filling in the ASA is achieved by
inserting empty spheres and by inflating the atom-centered nonoverlapping
spheres. The atomic radii are chosen in such a way that (i) the charge on
the empty spheres is negligible and (ii) the overlap of the interstitial
with the interstitial, atomic with the atomic, and interstitial with the
atomic spheres remains within the permissible limit of the ASA. The basis
set for the self-consistent electronic structure calculation for Ba$_{2}$%
Cu(PO$_{4}$)$_{2}$ includes Cu($s$, $p$, $d$), Ba($s$, $d$ and $f$), O($s$, $p$), and P($s$%
, $p$), and for Sr$_{2}$Cu(PO$_{4}$)$_{2}$ the basis set includes Cu($s$, $p$%
, $d$), Sr($s$, $d$), O($s$, $p$), and P($s$, $p$), the rest being downfolded. The
(8, 8, 8) $k$ mesh has been used for self-consistency. All the $k$-space
integrations were performed using the tetrahedron method. \cite{tetra} In
order to extract the various hopping integrals we have employed NMTO based
downfolding method. \cite{nm1,nm2,nm3} The downfolding method consists of deriving a
few-orbital effective Hamiltonian from the full LDA Hamiltonian by
integrating out high-energy degrees of freedom.

\section{Results and Discussion}

The all-orbital band structure and the total density of states (DOS) and partial DOS for Ba$%
_{2}$Cu(PO$_{4}$)$_{2}$ and Sr$_{2}$Cu(PO$_{4}$)$_{2}$ are displayed in Fig.
2. The bands are plotted along the various high symmetry points of the
Brillouin zone corresponding to the monoclinic lattice. All the energies are
measured with respect to the Fermi level of the compound. The characteristic
feature of the non-spin-polarized
band structure displayed in Figs. 2(a) and 2(d) is an isolated
half-filled band at the Fermi level. This band is predominantly derived from
the antibonding linear combination of Cu $d_{x^{2}-y^{2}}$ and O $p_{\sigma }
$ states residing in the same square plaquette. The conduction band is
separated from the other Cu $d$ character dominated valence bands by a small
gap. This is consistent with the schematic energy level diagram displayed in Fig.
1(c). Further below these Cu $d$ bands, are the bands with
predominantly oxygen 2$p$ character. The conduction band disperses only
along the chain direction with hardly any dispersion perpendicular to the
chains. This is reflected in the partial DOS shown in the inset of Figs. 2b and 2e,
which has a characteristic feature of 1D tight-binding dispersion.
\begin{figure}[tbp]
\includegraphics{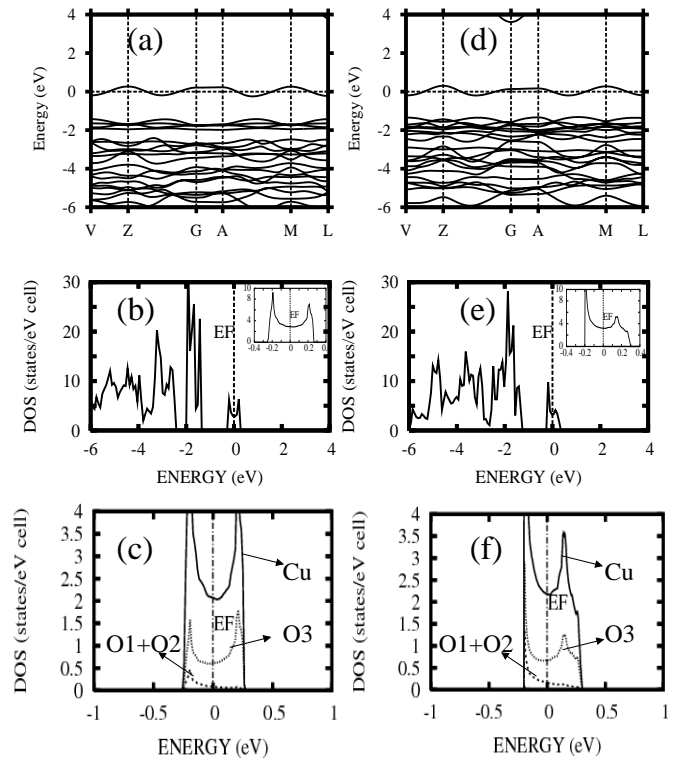}% Here is how to import EPS art
\caption{Band structure, density of states, and partial DOS of [(a)-(c)] Ba$_2$Cu(PO$_4$)$_2$ and [(d)-(f)] Sr$_2$Cu(PO$_4$)$_2$.}
\label{fig:banddos}
\end{figure}

This isolated Cu $d_{x^{2}-y^{2}}$ conduction band is responsible for the
low-energy physics of this material. This band hardly couples with any other
states except for the oxygens (O3) residing in the square plaquette as can
be seen from the plot of the oxygen partial density of states displayed in
Figs. 2(c) and 2(f). Since the calculations are carried out in the framework of LDA, the
system is not insulating. We have checked that the insulating behavior can
be recovered by including the Coulomb interaction in the framework of LDA +
$U$. In the following we shall employ NMTO downfolding method to map our LDA
results to a low-energy orthogonal tight-binding Hamiltonian. This
Hamiltonian will serve as the single electron part of the correlated Hubbard
model relevant for this system, which in turn can be mapped to an extended
Heisenberg model [Eqn. (1)] with the exchange couplings related to the LDA
hoppings by the relation 
\begin{equation}
J_{ij}=\frac{4t_{ij}^{2}}{U_{eff}}
\end{equation}
where $U_{eff}$ is the screened Coulomb interaction.

The NMTO downfolding method provides an \textit{ab initio} scheme to
construct a low-energy tight-binding Hamiltonian starting from the full
all-orbital LDA calculations. This method relies on the energy-selective
downfolding process to integrate out high-energy degrees of freedom
resulting in a low-energy Hamiltonian defined in the basis of effective
orbitals. These effective orbitals, by construction, are tailored to contain
in their tails the states which are integrated out with weights being
proportional to their admixture with the orbitals that are retained in the
basis. If the low-energy set of bands are separated from all other bands,
the orthonormalized NMTO set converges to a set of Wannier functions. Hence,
with the NMTO downfolding method, we can generate localized Wannier
functions directly without recourse to Bloch functions. These effective
orbitals (NMTOs) provide a direct visualization of the various interaction
paths. The Fourier transform of this few orbital downfolded Hamiltonian
provides various hopping integrals t$_{ij}$ between these effective orbitals,
and the corresponding tight-binding Hamiltonian can be written as

\begin{equation}
H_{TB} = \sum_{(ij)}t_{ij}(c_i^\dagger c_j + {\bf H}.\text{c}.),
\end{equation}

where $i$ and $j$ denote a pair of orbitals. These hopping integrals thus
form the first principles set of parameters obtained without any fitting
procedure containing the signature of the pathways involved in the hopping
process.

For the present compounds with one predominantly Cu $d_{x^{2}-y^{2}}$ band
so well separated from the rest, one can have a tight-binding model with a
single Cu $d_{x^{2}-y^{2}}$ orbital per Cu to be a good approximation to the
full band structure close to the Fermi level. In Fig. 3, we have displayed
the band structure of Ba$_{2}$Cu(PO$_{4}$)$_{2}$ and Sr$_{2}$Cu(PO$_{4}$)$%
_{2}$ decorated with Cu $d_{x^{2}-y^{2}}$ character together with the
downfolded band structure where only Cu $d_{x^{2}-y^{2}}$ orbital is
retained in the basis and the rest are downfolded. The agreement between the two
is remarkable, suggesting that our one band Hamiltonian is indeed physically
reasonable and is suitable to capture the low-energy physics of the system.
The Fourier transform of the downfolded Hamiltonian $H(k)\longrightarrow H(R)
$ gives the effective hopping parameters for the physical Hamiltonian. In
Table I, we have displayed the various dominant effective hopping integrals $%
t_{ij}$ (having magnitude $\geq $ 1 meV) between the Cu$^{2+}$ ions at sites 
$i$ and $j$. The notations for various hoppings for a pair of Cu 
site ($i$ and $j$) are indicated in Fig. 1(b), where, e.g., for a Cu ion at 
site $i$, $t_{ij}=t_1$ if the other Cu ion at site $j$ is at the nearest 
neighbor-position along the chain and similarly for all the other hoppings.

From Table I, we gather that the nearest-neighbor hopping (t$_{1}$) 
is dominant for
both the compounds. The other hoppings (t$_{2}$) 
along the chain (intrachain) as well
as between the chains (interchain) ($t_{1}^{ic}$, $t_{2}^{ic}$) and between
chains in the perpendicular ($c$) direction ($t_{\bot}$) are an order of magnitude smaller than
the dominant nearest-neighbor hopping. The smaller magnitude of the hopping
for Ba$_{2}$Cu(PO$_{4}$)$_{2}$ may be attributed to the relatively large
lattice constant for the compound. Recently, these hoppings were also
calculated by adopting a fitting procedure. \cite{ros} Although the trends
among the hoppings obtained by fitting the conduction band agree with our 
\emph{ab initio} results, the magnitude as well as the range of the
hoppings are exaggerated in the fitting scheme. 
\begin{figure}[tbp]
\includegraphics{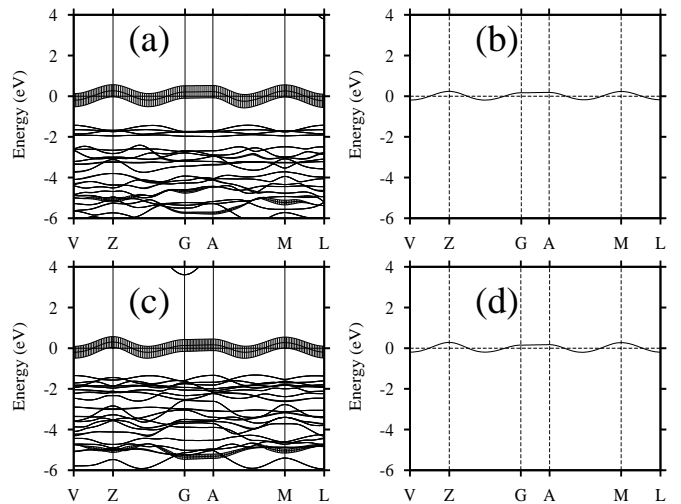}% Here is how to import EPS art
\caption{Cu $d_{x^2-y^2}$ projected as well as downfolded band in [(a) and
(b)] Ba$_{2}$Cu(PO$_{4}$)$_{2}$ and [(c) and (d)] Sr$_{2}$Cu(PO$_{4}$)$_{2}$.}
\label{fig:rt}
\end{figure}

In Fig. 4, we have displayed
the Cu $d_{x^{2}-y^{2}}$ Wannier function. We note that for both 
the compounds, the Cu $%
d_{x^{2}-y^{2}}$ Wannier function is primarily localized in one plaquette
and it hardly couples with the neighboring plaquettes. The tails of the Cu $%
d_{x^{2}-y^{2}}$ orbital are shaped according to O $p_{x}/p_{y}$ orbitals
such that the Cu $d_{x^{2}-y^{2}}$ orbital forms strong pd$\sigma $
antibonds with the O $p_{x}/p_{y}$ tails. We also note that the Cu $%
d_{x^{2}-y^{2}}$ Wannier function for Ba$_{2}$Cu(PO$_{4}$)$_{2}$ is more
localized, explaining the smaller magnitude of the hoppings for this system.

\begin{figure}
\includegraphics{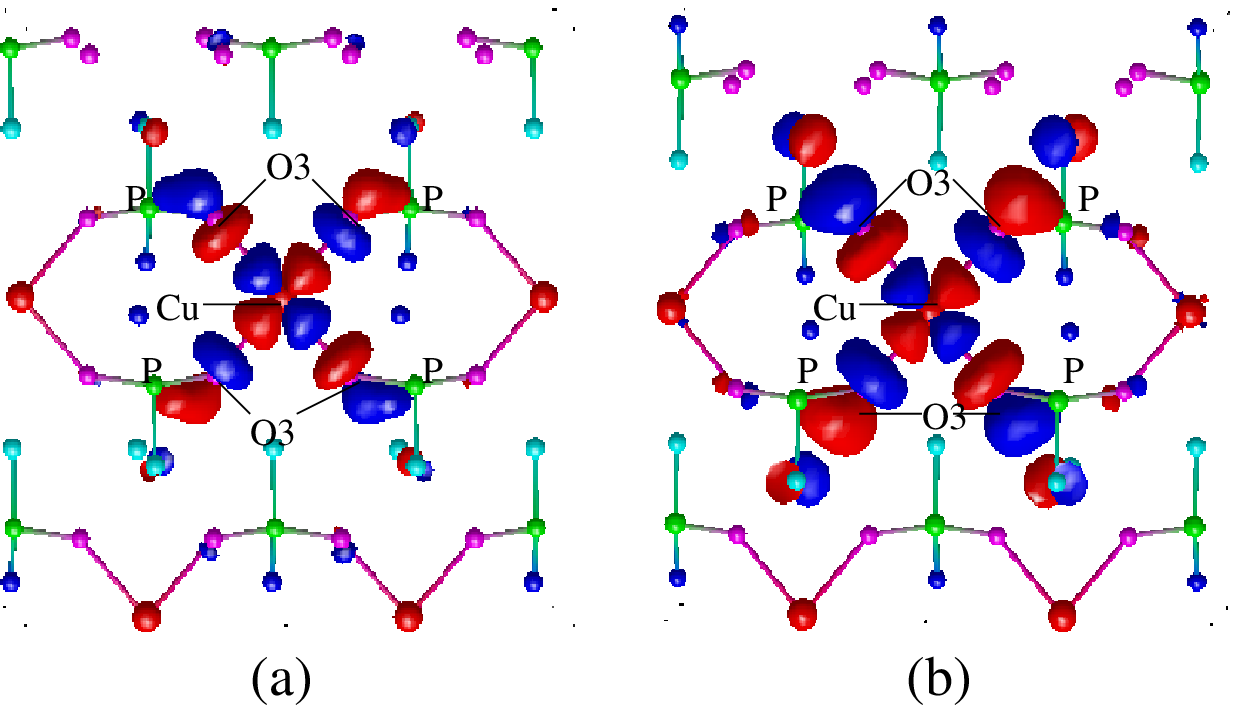}% Here is how to import EPS art
\caption{(Color online) 
\label{fig:rt} Effective $Cu-d_{x^2-y^2}$ Wannier function plot for 
(a) Ba$_{2}$Cu(PO$_{4}$)$_{2}$ and (b) Sr$_{2}$Cu(PO$_{4}$)$_{2}$.
The $d_{x^{2}-y^{2}}$ orbital is defined with the choice of local 
co-ordinate system as discussed in the text. The spheres represent 
the ions as indicated in the figure.}
\end{figure}

The hopping parameters listed in Table I can be used to estimate various
exchange interactions using Eq.(2). In the absence of a
satisfactory way of direct computation of exchange integrals, such an
approximate method is a good starting point for estimating the exchange
couplings as well as the relative strengths of the various exchange
integrals. The ratios of the various exchange interactions for each compound
are listed in Table II. As expected, the nearest-neighbor exchange interaction
is dominant for both the compounds with practically negligible interchain
and further-neighbor intrachain interactions, making these compounds 
ideal 1D HAFs spanning a wide temperature range. The agreement between the
ratio of the nearest-neighbor exchange interaction in Ba$_{2}$Cu(PO$_{4}$)$
_{2}$ to that in Sr$_{2}$Cu(PO$_{4}$)$_{2}$,  $\frac{J_{Ba}}{J_{Sr}}$,
obtained in our calculation and that obtained from experiment is remarkable.
In the following, we shall try to understand the origin of the 
short-range hopping and
therefore the exchange interactions in these systems. From the Cu $
d_{x^{2}-y^{2}}$ Wannier function (Fig. 4), we gather that the Cu-Cu hopping
primarily proceeds via the oxygens.
As a result, as argued by Koo {\em et. al.} \cite{whangbo} 
the strength of the Cu-O...Cu spin-exchange is
primarily governed by the O...O distance and Cu-O..Cu angles rather than the
Cu..Cu distances. The exchange interaction therefore becomes negligible when
the relative separation between the oxygens mediating the interaction is
large compared to the van der Waals distance. For the present compounds, the
weak hopping integrals and therefore the exchange interactions between the Cu
ions may be attributed to their peculiar geometry. The CuO$_{4}$ plaquettes
in the present compounds neither share edges nor corners and are connected
by PO$_{4}$ tetrahedra resulting in a super-super exchange process to
mediate the exchange interactions. This arrangement results in a large
relative separation (greater than the van der Waals distance) between the oxygens except
for the oxygens in the neighboring plaquettes along the chain, resulting in an appreciable
nearest-neighbor interaction while other intrachain as well as interchain
interactions are negligible.

As mentioned earlier, even tiny 
interchain interactions can cause a deviation from
the ideal 1D behavior. \ Indeed, in the $^{31}$P NMR shift (which is
proportional to the spin susceptibility) measurements on Ba$_{2}$Cu(PO$_{4}$)%
$_{2}$ and Sr$_{2}$Cu(PO$_{4}$)$_{2}$, \cite{nath} it was found that while
the data are well described by the 1D HAF model in a large temperature
range, a sharp deviation was seen below $k_{B}T$/$J\approx $ $0.03$ (still
above the LRO\ temperature found by Belik \textit{et al.} \cite{belik}). \ A
small interchain interaction can therefore be of significance at very low
temperatures. \ \ In order to check that small interchain hoppings are
indeed important, we calculated the zero-temperature spin-spin correlation
function for a pair of staggered $S=\frac{1}{2}$ Heisenberg chains [see Fig. 1(b)]
(24 sites, with periodic boundary conditions) using
the exact diagonalization method. We confirmed that in the absence of
interchain coupling, the spin-spin correlation function decays with a
characteristic power-law behavior. However, inclusion of a very tiny
interchain coupling $\frac{J_{ic}}{J_{1}}$ $\approx$ 0.02
causes the spin-spin correlation function to deviate from
the power-law behavior and acquire an exponential decay. \ This might then
be responsible for the sharp decrease in the $^{31}$P NMR shift seen at
low-temperatures by Nath \textit{et al}. \cite{nath}

\section{Summary and Conclusions}

We have employed TB-LMTO ASA and NMTO method to analyse in detail the 
electronic structure of two isostructural phosphates, Ba$_{2}$Cu(PO$_{4}$)%
$_{2}$ and Sr$_{2}$Cu(PO$_{4}$)$_{2}$. The characteristic feature 
of the LDA electronic structure of these compounds is an isolated 
half-filled band at the Fermi level derived from the antibonding linear 
combination of Cu $d_{x^{2}-y^{2}}$  and oxygen $p_{\sigma}$ states
residing on CuO$_{4}$ square plaquettes. The various hopping integrals 
obtained using NMTO downfolding method suggest that 
these compounds are indeed one dimensional with dominant nearest-neighbor
hopping integrals along the chains. Our estimate of the exchange interaction,
in particular the ratio of the exchange interaction of 
Sr$_{2}$Cu(PO$_{4}$)$_{2}$ to that of Ba$_{2}$Cu(PO$_{4}$)$_{2}$, compares 
excellently with the experimental results. The plot of the Cu $d_{x^{2}-y^{2}}$
Wannier functions  suggest that the exchange interactions between 
the magnetic Cu$^{2+}$ ions is primarily mediated by the oxygens.
Our calculations confirm that the 
unique geometry of the CuO$_{4}$ plaquettes, which are neither face sharing nor 
corner sharing, is responsible for the unique 1D magnetic properties 
for these systems. Finally, we argue that although the interchain couplings are small, they may still be important for the low-temperature properties for this 
system and may explain the sharp deviation from the 1D HAF model 
at low-temperature seen in the $^{31}$P NMR shift for these systems.

\begin{table}[tbp]
\caption{Hopping integrals obtained from our ab initio analysis for Ba$_2$%
Cu(PO$_4$)$_2$ \& Sr$_2$Cu(PO$_4$)$_2$.}
\label{tab:table1}
\begin{ruledtabular}
\begin{tabular}{cccccccccc}
 & t$_1$ (meV) &t$_2$ (meV)& t$_1$$^{ic}$ & t$_2$$^{ic}$ & t$_{\bot}$   \\
\hline
Ba$_2$Cu(PO$_4$)$_2$& 96 & 4 & 5 &1
& 2 \\
Sr$_2$Cu(PO$_4$)$_2$& 103 & 9 & 14 &0
& 1   \\
\end{tabular}
\end{ruledtabular}

\end{table}

\begin{table}[tbp]
\caption{Ratios of the exchange coupling constants and comparison with experiment for Ba$_2$%
Cu(PO$_4$)$_2$ \& Sr$_2$Cu(PO$_4$)$_2$.}
\label{tab:table2}
\begin{ruledtabular}
\begin{tabular}{cccccccccc}
 &\it{J}$_2$/\it{J}$_1$  &\it{J$_1$}$_{ic}$/\it{J}$_1$ & \it{J}$_{Ba}$/\it{J}$_{Sr}$\footnotemark[1]& \it{J}$_{Ba}$/\it{J}$_{Sr}$\footnotemark[2]  \\
\hline
Ba$_2$Cu(PO$_4$)$_2$& 0.0017 & 0.0027 & 0.869& 0.915\\
Sr$_2$Cu(PO$_4$)$_2$& 0.0076 & 0.0184  \\
\end{tabular}
\end{ruledtabular}
\footnotetext[1]{Ratio of $\it{J}$/$\it{K}$$_{B}$ of Ba$_2$Cu(PO$_4$)$_2$ to that of Sr$_2$Cu(PO$_4$)$_2$ obtained from theory.}
\footnotetext[2]{Ratio of $\it{J}$/$\it{K}$$_{B}$ of Ba$_2$Cu(PO$_4$)$_2$ to that of Sr$_2$Cu(PO$_4$)$_2$ obtained from experiment. \cite{nath}}

\end{table}

\begin{acknowledgments}
I.D. thanks the DST, India (Project No. SR/S2/CMP-19/2004) for financial support. A.V.M. thanks Alexander von
Humboldt foundation for the financial support for the stay at Augsburg.
\end{acknowledgments}
\bibliography{apssamp}
% Produces the bibliography via BibTeX.

\end{document}